\documentclass[a4paper, 12pt]{article}

\usepackage{amsmath,amsfonts,amssymb,epsfig,psfrag}
\usepackage{subfigure}
\usepackage{amsmath,amsfonts,amssymb}
\usepackage{graphicx, color}

\newcommand{\bR}{\pmb{{\cal R}}}

\renewcommand{\ge}{\geqslant}

\newcommand{\demi}{\textstyle{\frac{1}{2}}}

\renewcommand{\vec}[1]{\boldsymbol{#1}}
\begin{document}
\numberwithin{equation}{section}

\title{Large amplitude Love waves}
\author{E. RODRIGUES FERREIRA, Ph. BOULANGER, \\ M. DESTRADE}

\date{2008}

\maketitle

%++++++++++++++++++++

\begin{abstract}

In the context of the finite elasticity theory we consider a model for
compressible solids called ``compressible neo-Hookean material''.
We show how finite-amplitude inhomogeneous plane wave solutions
and finite-amplitude unattenuated solutions can combine to form a
finite-amplitude Love wave.
We take a layer of finite thickness overlying a solid half-space,
both made of different pre-stressed compressible neo-Hookean materials.
We derive an exact solution of the equations of motion and boundary
conditions, and
also obtain results for the energy density and the energy flux of the waves.
Finally, we investigate the special case when the interface between the layer
and the substrate is in a principal plane of the pre-stain. A
numerical example is given.

\end{abstract}

%++++++++++++++++++++++++++++++

\section{Introduction}

A seismic event launches at
least two types of \emph{surface waves}, one causing vertical
(elliptic) movements, the other causing rather destructive lateral
(shear horizontal) movements. In the essay which won him the Adams
Prize in 1911, Love \cite{Love11}
proposed a simple Earth model which supports the latter kind of waves,
by considering a crust made of an isotropic, linear, elastic solid,
rigidly bonded onto a substrate (a semi-infinite solid) made of
another isotropic, linear, elastic solid.
In this heterogeneous structure, a shear horizontal wave may
propagate, leaving the upper face of the layer free of traction, and
having an amplitude which decays rapidly with depth in the substrate.
This localisation of the amplitude variation is what makes Love waves
(and surface waves in general) a subject of great interest in
seismology, because it means that the energy spreads essentially in
two dimensions, and thus the wave travels further from the epicentre
than bulk waves,  for which the energy spreads in three dimensions.

Over the years, Love's results were extended in several directions 
and turned
out to be also useful in other contexts. For instance 
anisotropy (e.g. \cite{LaMT71}),
inhomogeneity  (e.g. \cite{Will42, Dere62, Bhat70}), piezoelectric 
coupling (e.g.
\cite{Bleu68, Guly69, CoDe06}), and many other effects
were considered \textemdash see the review by Maugin \cite{Maug83} for 
an exhaustive account.
Two possibilities are of special interest to seismology science:
the possibility of including \emph{strain-induced anisotropy}, because
it provides a simple and revealing modelling of the consequence of
slow tectonic movements, and the possibility of including
\emph{non-linear effects}, because large amplitude seismic movements
have indeed been observed. The first possibility can be dealt 
 with within
the framework of small-amplitude waves superimposed upon a large static
homogeneous pre-strain, see Hayes and Rivlin \cite{HaRi61}, Willson
\cite{Will75}, Kar and Pal \cite{KaPa85}, or Dowaikh \cite{Dowa99}.
The second possibility is usually treated in the framework of the
so-called weakly non-linear elasticity theory (see Maugin
\cite{Maug99} or Norris \cite{Norr99} for instance), where the
equations of motion are developed one order further than the linear
regime (Note that according to Zabolotskaya \cite{Zabo86}, non-linear
shear horizontal waves require fourth-order elasticity.)

Here we propose to combine both effects, by considering
finite-amplitude Love waves in a finitely deformed layer/substrate structure.
Our only restriction is in our choice of a constitutive law for the
solids, as we focus on the so-called \emph{compressible neo-Hookean
materials}.  These enjoy a peculiar property once deformed \cite{1, 2, 3}:
they allow the propagation of a finite-amplitude, inhomogeneous,
linearly polarized, transverse plane wave in any direction.
Moreover,   this wave is obtained by solving 
a linear ordinary
differential equation even though the theory is completely non-linear (In
passing we note that the eventuality of a solitary wave is thereby
precluded here). We use this wave as  an ingredient in  
the construction of the Love wave
solution to the corresponding boundary value problem.
A great deal of generality is nonetheless achieved, in particular
because \emph{non-principal} wave propagation is possible, and because
the solution is \emph{exact}, without any limitations to be imposed
on its magnitude.
Corresponding general results are obtained on energy propagation in
the layer and the substrate.

%Of course we remain aware that no `real' solid is perfectly described by the
%`compressible neo-Hookean' strain-energy density.
%In fact, some calculations (not reproduced here) conducted on the 
%Mooney-Rivlin
%strain-energy density, which is slightly more complicated, indicate that it is
%not possible to construct the same of solution for other solids.
%In all generality, a finite amplitude surface wave will probably not 
%be linearly
%(Love wave) or elliptically (Rayleigh wave) polarized, and its motion
%will be governed by a nonlinear equation.
%We argue that the exact and explicit results of this paper will then provide
%a safe a reliable benchmark for other strain-energy densities.

A special case of the `compressible neo-Hookean' model was first 
introduced to describe a
class of solid polyurethane rubbers studied in the Blatz-Ko 
experiments \cite{BlatzKo}.
Of course we remain aware that solids playing a role in the 
applications of Love waves are
not always adequately described 
 by the `compressible neo-Hookean' strain energy 
density. However
exact and expilicit results like those of this paper are always 
useful, either as a basis
for perturbation methods or for testing numerical schemes. 

The paper is organised as follows. In Section \ref{section2},
we present the
constitutive equations of the
compressible neo-Hookean model. Next
(Section \ref{section3}), we   retrieve 
results for transverse inhomogeneous time-harmonic waves superimposed
on a pre-stressed state \cite{1, 2} but also  
present similar results for transverse unattenuated time-harmonic motion.
In Section \ref{section4}, the set-up consisting of a
semi-infinite substrate
covered with a layer of finite thickness is described and the effect of
pre-strain is investigated, assuming that the two solids are rigidly bonded.
A large amplitude Love wave solution is then obtained provided the propagation
direction in the interface and the normal to the interface are along conjugate
directions of the pre-strain tensors. A dispersion equation is also
obtained, similar to that
of linear isotropic elasticity. Here however this equation involves the bulk
wave speeds along the propagation direction and along the normal to
the layer, which are
both affected by the pre-deformations. In Section \ref{section6}, we derive the
energy densities and
energy fluxes corresponding to the motions in the layer and in the
substrate.
Mean energy densities and fluxes are obtained by averaging over a
period in time, and
their properties are investigated.
Total energy densities and
fluxes are also
introduced, showing that the repartition of energy between the layer
and the substrate
depends on the ratio of layer thickness to the wavelength.
Finally (Section \ref{section7}), we consider the special case when
the interface is in a
principal plane of the pre-strain tensors in the layer and the
substrate. In this case the
propagation direction may be any direction in the interface and,
owing to the anisotropy
induced by the pre-stress, the Love wave speed varies with the
propagation direction.

%++++++++++++++++++++++++++++++

\section{Compressible neo-Hookean materials}
\label{section2}

%++++++++++++++++++++++++++++++

In order to deal with possibly large deformations of solids,
we invoke the finite elasticity theory.
To model the non-linear elasticity of solids, we use one of the simplest
constitutive models on offer for compressible isotropic solids, which
may be called  the  `compressible neo-Hookean 
material'.
  We here present this model.

Let $\vec{F}$ be the \emph{deformation gradient}, defined as usual
(see, for instance, \cite{Chad99}) by 
\begin{equation}
\vec{F}= \partial \vec{x} / \partial \vec{X},
\qquad F_{iA}=\partial x_i /\partial X_A,
\end{equation}
 where $\vec{X}$ is the position of a particle in the 
reference (Lagrangian)
configuration and $\vec{x}$ the corresponding position in the current 
(Eulerian)
configuration.  
Associated with $\vec{F}$ is the the left Cauchy-Green strain tensor
\begin{equation} \label{B}
     \vec{B} = \vec{F} \vec{F}^T,
\qquad B_{ij}=(\partial x_i /\partial X_A) (\partial x_j /\partial X_A),
\end{equation}
where the superscript $T$ denotes the transpose.
The determinant of $\vec{F}$ is
\begin{equation}
J = \det \vec{F} =(\det \vec{B})^{1/2}.
\end{equation}
It is the ratio between the volume of a material element of the solid
in the reference and current states.
Then, \emph{compressible neo-Hookean materials} are
characterized by a strain-energy density $W$, measured per unit volume in the
undeformed state, given by
\begin{equation}
2W=\mu (\mbox{tr}\, \vec{B}\ -3)+G(J)-G(1),
\label{W}
\end{equation}
where $\mu$ is a constant (the \emph{shear modulus})
and $G(J)$ is an \emph{arbitrary} function of $J$ (a material function,
which can be adjusted to model the compressibility properties of the solid).
The corresponding constitutive equation for the symmetric
Cauchy stress tensor
$\vec{T}$ is
\begin{equation} \label{eq:BlatzKo}
          \vec{T} = \demi G'(J)\vec{I} + \mu J ^{-1}\vec{B}.
\end{equation}
Such a `compressible neo-Hookean' material is sometimes called
`special Blatz-Ko' material,  or `restricted Hadamard'
material \cite{1, 2, 3}.
Hayes \cite{Haye68} shows that the resulting equations of motion
are strongly elliptic when
\begin{equation} \label{S-E}
\mu >0\ , \qquad G''(J) \ge 0,
\end{equation}
and we assume as much henceforth.
We also assume that the undeformed state is stress free, so that
$G'(1)=-2\mu$.
Note that comparison with linearised isotropic elasticity yields
$G''(1) = 2(\lambda + \mu)$, where $\lambda$ and $\mu$ are the Lam\'e
coefficients.
Here however, no restriction is placed on the amplitude of the
displacement $\vec{u}=\vec{x}-\vec{X}$, so that due to \eqref{B},
and the arbitrariness of $G(J)$, the relation between
$\vec{T}$ and $\vec{u}$ is clearly non-linear.

Several examples of specific volumetric functions $G(J)$ have
been presented over the years (see, for instance, Ba\c{s}ar
and Weichert \cite{BaWe00}). Among these, we recall the
particularly simple
choice of Levinson  and  Burgess  \cite{LeBu71}, leading to a model called
`simplified Blatz-Ko' in \cite{3}, that is
\begin{equation}
G(J) = (\lambda + \mu)(J^2-1) - 2(\lambda + 2 \mu)(J-1),
\label{LevBurg}
\end{equation}
with $\mu > 0$ and $\lambda + \mu \geq 0$ in order to satisfy
\eqref{S-E}.

%++++++++++++++++++++++++++++++++++++++++++++++++++++++++++++++

\section{Transverse waves superimposed on a static deformation}
\label{section3}

%++++++++++++++++++++++++++++++++++++++++++++++++++++++++++++++

Here we consider transverse wave solutions in unbounded pre-strained materials.
Suppose that a compressible neo-Hookean material is first subjected to a
static finite homogeneous deformation
defined by
\begin{equation}
\vec{x} = \vec{F} \vec{X}, \quad x_i = F_{iA} X_A,
\label{eq:3.1}
\end{equation}
where the $F_{iA}$ are constants. The corresponding
constant left Cauchy-Green strain tensor is
$\vec{B}=\vec{F}\vec{F}^T$ and the determinant $J = \text{det }\vec{F}$
is a constant. On this state of deformation, we superpose a
time-dependent displacement
taking a particle from position $\vec{x}$ to position
\begin{equation}
           \overline{\vec{x}}=\overline{\vec{x}}(\vec{x}, t)=
	\overline{\vec{x}}(\vec{F}\vec{X}, t)=
	\vec{x} + \vec{u}(\vec{x}, t), \label{eq:displace}
\end{equation}
where $\vec{u}$ is the mechanical displacement.
In the absence of body forces,
the equations of motion for this time-dependent deformation may be
written in the form
\cite{1, 3}
\begin{equation}
\rho\, \ddot{\overline{\vec{x}}}= {\rm div}_{\vec{x}}\overline{\vec{P}} ,
\qquad  \rho\, \ddot{\overline{x}}_i = \partial \overline{P}_{ik} /
\partial x_k,
\label{eq:3.17}
\end{equation}
where $\rho =\rho_0 J^{-1}$ is the constant mass density in the
intermediate state of static deformation, and $\overline{\vec P}$
is the Piola-Kirchhoff
stress tensor at time $t$ with respect to the intermediate state of 
static deformation,
\begin{equation}
\overline{\vec{P}}=(\det
\widehat{\vec{F}})\overline{\vec{T}}\widehat{\vec{F}}^{-T} ,
\qquad \overline{P}_{ik}=(\det
\widehat{\vec{F}})\overline{T}_{ij}\widehat{F}_{kj}^{-1} .
\label{eq:3.16}
\end{equation}
Here $\overline{\vec T}$ is the Cauchy stress tensor at time $t$, and
$\widehat{\vec{F}} = \partial \overline{\vec{x}}/\partial \vec{x}$ is
the deformation
gradient with respect to the
intermediate state of static
deformation. We note that
$\widehat{\vec{F}}=\overline{\vec{F}}\vec{F}^{-1}$, where
$\overline{\vec{F}} = \partial \overline{\vec{x}}/\partial \vec{X}$
is the deformation
gradient with respect to the undeformed
configuration.

 As in \cite{2}, we are now looking for 
solutions with a
displacement field $\vec{u}=\overline{\vec{x}}-\vec{x}$ of the form
\begin{equation}
\vec{u} =  f(\vec{m} \cdot \vec{x}) g(\vec{n} \cdot {\bf
x}-vt)\vec{a} , \label{eq:3.3}
\end{equation}
where $f$ and $g$ are functions to be determined, and $\vec{m}$, $\vec{n}$ and
$\vec{a}$ are unit vectors. It is assumed that $\vec{m}$ and
$\vec{n}$ are not parallel and that
$\vec{a}$ is orthogonal to both $\vec{m}$ and $\vec{n}$,
\begin{equation}
\vec{a}\cdot \vec{m} =\vec{a}\cdot \vec{n} = 0,
\label{eq:3.4}
\end{equation}
so that (\ref{eq:3.3}) represents a linearly polarized transverse
wave with propagation speed $v$.

For such a wave motion, \color{black} recall that \color{black} 
$\widehat{\vec{F}}$
and its inverse are \cite{2}
\begin{equation} \label{Fhat}
\widehat{\vec{F}} =
      \vec{I} + f' g \vec{a} \otimes \vec{m} + f g' \vec{a} \otimes \vec{n},
\qquad
\widehat{\vec{F}}^{-1} =
      \vec{I} - f' g \vec{a} \otimes \vec{m} - f g' \vec{a} \otimes \vec{n},
\end{equation}
\color{black} so that \color{black} the special Blatz-Ko constitutive equation
(\ref{eq:BlatzKo}) \color{black} yields \color{black}
\begin{equation} \label{eq:Tbar}
\overline{\vec{T}} =
     \demi G'(J) \vec{I} + \mu J^{-1} \widehat{\vec{F}}
        \vec{B} \widehat{\vec{F}}^T,
\end{equation}
and, because $\det \widehat{\vec{F}}=1$, the Piola-Kirchhoff stress 
tensor \eqref{eq:3.16}
reduces here to
\begin{equation} \label{Pbar}
\overline{\vec{P}} = \demi G'(J) \widehat{\vec{F}}^{-T}
     + \mu J^{-1} \widehat{\vec{F}} \vec{B}.
\end{equation}

\color{black} It follows \cite{2} \color{black}
that the displacement field (\ref{eq:3.3}) is a solution of
the equations of motion if and only if $f$ and $g$ satisfy the equation
\begin{equation}
(\vec{n} \cdot\vec{Bn}- \mu^{-1}\rho_0 v^2)fg''+
2\vec{n} \cdot\vec{Bm}f'g'+\vec{m} \cdot\vec{Bm}f''g = 0,
        \label{eq:3.19}
\end{equation}
where $f'$ and $g'$ denote the derivatives of $f$ and $g$ with
respect to their argument.
We then choose $\vec{m}$ and $\vec{n}$ such that  \cite{1}
\begin{equation}
       \vec{n} \cdot\vec{B} \vec{m}=0, \label{nBm}
\end{equation}
which means that $\vec{m}$ and $\vec{n}$ are along the principal
axes fo the elliptical section of the ellipsoid $\vec{x \cdot B x} = 1$
by the plane $\vec{a \cdot x} = 0$. Then, (\ref{eq:3.19}) yields two uncoupled
equations for $f$ and $g$~:
\begin{equation}
         v^{2}_{\vec{m}} (f'' / f) = -c,  \qquad
         (v^{2}_{\vec{n}}-v^2) (g'' / g) = c \label{eqfg} ,
\end{equation}
where $c$ is an arbitrary constant, and $v_{\vec{m}}$ and
$v_{\vec{n}}$ are the wave speeds
of homogeneous bulk waves propagating along $\vec{m}$ and $\vec{n}$,
respectively,
\begin{equation} \label{homogv}
\rho_{0}v^{2}_{\vec{m}} = \mu\, \vec{m}  \cdot\vec{B}  \vec{m}  , \qquad
\rho_{0}v^{2}_{\vec{n}} = \mu\, \vec{n}  \cdot\vec{B}  \vec{n}.
\end{equation}

If $c$ is assumed to be \emph{positive}, $c=\kappa^2$ (say) for some 
real $\kappa$, then
(\ref{eqfg}) yields an \emph{unattenuated
time-harmonic wave motion} provided $v^2 > v^{2}_{\vec{n}}$. The
displacement field of this
wave is
\begin{multline} \label{Unatt}
\vec{u} (\vec{x}, t) = \left[B\sin(\frac{\kappa}{v_{ \vec{m}}}
\vec{m} \cdot \vec{x}) + C\cos(\frac{\kappa}{v_{\vec{m}}} \vec{m} \cdot
\vec{x})\right]\\ \times
\cos\left(\frac{\kappa}{\sqrt{v^2-v^{2}_{\vec{n}}}} (\vec{n} \cdot
\vec{x}-vt)\right)\vec{a},
\end{multline}
where $\kappa$, $B$, $C$ are arbitrary constants.

If $c$ is assumed to be \emph{negative}, $c=-\gamma^2$ (say) for some 
real $\gamma$, then
(\ref{eqfg}) yields an
\emph{inhomogeneous time-harmonic wave motion} provided $v^2 <
v^{2}_{\vec{n}}$.
The displacement field of this inhomogeneous plane wave is
\begin{equation} \label{Att}
\vec{u}  (\vec{x}, t) = A \exp\left(-\frac{\gamma}{v_{ 
\vec{m}}}\vec{m}\cdot \vec{x}\right)
\cos \left(\frac{\gamma}{\sqrt{v^{2}_{
\vec{n}}-v^2}}(\vec{n}\cdot\vec{x}-vt) \right)\vec{a},
\end{equation}
where $\gamma$, $A$ are arbitrary constants. \color{black} Here we 
retrieve a solution
obtained in \cite{1, 2}. However the solution \eqref{Unatt} was 
not mentionned in
these papers because the emphasis
 \color{black} there \color{black}  was on inhomogeneous plane waves. 
 Here, both
\eqref{Unatt} and \eqref{Att} are needed for the construction of a 
Love wave solution.
\color{black}

We remark that when the condition (\ref{nBm}) is not satisfied,
solutions may nevertheless be
obtained \cite{2}; however either $f$ or $g$ is then of real exponential type
and hence no time-harmonic wave motion is possible.
In this paper we focus on time-harmonic waves
because they are the building blocks for Love waves.

For future reference, we conclude this section with
the evaluation of the traction vector $\vec{t}$ on a plane
$\vec{m}\cdot \vec{x}= \mbox{\it constant}\,$. Because the
displacement $\vec{u}$ is
along $\vec{a}$ and hence orthogonal to $\vec{m}$, such a plane is
globally preserved in
the motion. Moreover, $\det \widehat{\vec{F}}=1$  and
$\widehat{\vec{F}}^{-T}\vec{m}=\vec{m}$ by (\ref{Fhat}).
Hence, using Nanson's formula,
$\overline{\text{d}\vec{a}}=(\det
\widehat{\vec{F}})\widehat{\vec{F}}^{-T}\text{d}\vec{a}$, linking an
areal element $\overline{\text{d}\vec{a}}$ in the current configuration with
the same areal element $\text{d}\vec{a}$ in the intermediate state, we
conclude that when
$\text{d}\vec{a}$ is along $\vec{m}$, then the areal element is the 
same in both
configurations:
$\overline{\text{d}\vec{a}} = \text{d}\vec{a} = \vec{m} \text{d}a$. Thus, the
traction vector $\vec{t}$ on a plane $\vec{m}\cdot \vec{x}= \mbox{\it
constant}\,$ is the
same whether it is measured per unit area of the intermediate state
or per unit area of the
current state. Recalling (\ref{nBm}), we obtain
\begin{equation}
      \vec{t}=\overline{\vec{P}}\vec{m}=\overline{\vec{T}}\vec{m}=
      \vec{T m} + \rho v^2_{\vec m} f'g \vec{a},
      \label{eq:traction}
\end{equation}
where $\vec{T}$ denotes the constant Cauchy stress tensor of the
intermediate state.

%++++++++++++++++++++++++++++++

\section{The pre-stressed layered formation}
\label{section4}

%++++++++++++++++++++++++++++++

We wish to extend the classical results of Love \cite{Love11} in the linear
elasticity theory in two directions:
by taking account of initial stresses (and the accompanying
strain-induced anisotropy),
and by allowing the wave's amplitude to be arbitrarily large.

We start with Love's original set-up, which consists of a semi-infinite
substrate, covered with a layer of finite thickness.
The two solids are bonded rigidly. Here we assume that both the substrate
and the layer are made of different `compressible neo-Hookean materials',
with a shear modulus
$\mu$ and a function
$G$ for the substrate and a shear modulus $\widetilde{\mu}$ and a function
$\widetilde{G}$ for the layer. Also, $\rho_0$ and
$\widetilde{\rho}_0$ denote the
mass densities
of the substrate and the layer, respectively, measured in the undeformed
reference configuration.

In order to model geological formations, it is common to
consider that the solids
have been subjected to initial stresses, giving rise to
strain-induced anisotropy (see for instance the works of
Biot \cite{Biot63} or Tolstoy \cite{Tols73}).
To simplify matters here, we focus on static \emph{homogeneous
initial strains}. Thus, if $\vec{X}$ denotes the position
of a material particle in the undeformed solids, with origin $\vec{X}=\vec{0}$
in the interface between the substrate and the layer,
then the initial deformations are
\begin{equation}
      \vec{x}=\vec{FX}, \qquad
\widetilde{\vec{x}}=\widetilde{\vec{F}}\vec{X}.
      \label{eq:xFx}
\end{equation}
Here, the components of the deformation gradients
$\vec{F}$ in the substrate, and $\widetilde{\vec{F}}$ in the layer,
are constants.
The associated constant left Cauchy-Green strain tensors
are $\vec{B}=\vec{F}\vec{F}^T$ in the substrate and
$\widetilde{\vec{B}}=\widetilde{\vec{F}}\widetilde{\vec{F}}^T$ in the layer.
Also, we let $J =\det \vec{F}$ and $\widetilde{J} =\det \widetilde{\vec{F}}$.

\begin{figure}
\centering
\epsfig{figure=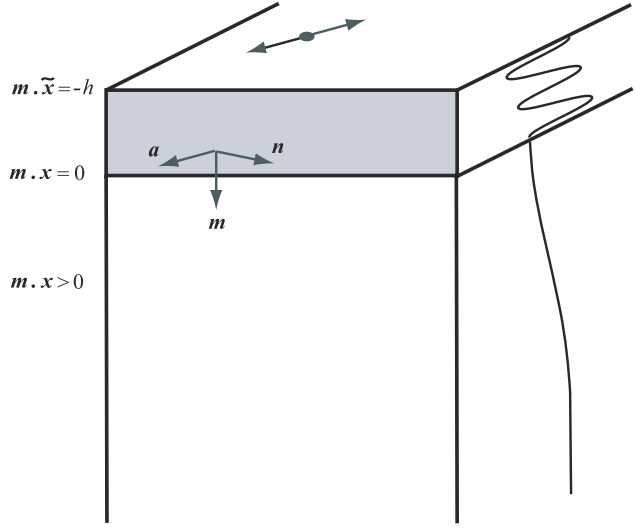, width=.6\textwidth}
   \caption{
   Sketch of the layered formation.
   The wave propagates in the direction of the unit vector $\vec{n}$,
   it is polarized along $\vec{a}$, and the magnitude of its amplitude varies
   along the direction of the unit vector $\vec{m}$, normal to the
faces of the layer.}
   \label{fig_sketch}
\end{figure}

We call $\vec{m}$ the unit vector normal to the faces of the layer,
in the static pre-strained state (\ref{eq:xFx}), oriented from the
layer toward the substrate, see Fig.\ref{fig_sketch}.
Hence, in this pre-strained state, the layer/substrate \emph{interface}
is the plane $\vec{m \cdot x} = 0$, or equivalently,
$\vec{m} \cdot \widetilde{\vec{x}} = 0$, and the substrate occupies
the $\vec{m \cdot x} \ge 0$ half-space. Also, in the static
pre-strained state, the
\emph{upper face} of the layer, in contact with vacuum, is the plane
$\vec{m} \cdot \widetilde{\vec{x}} = - h$, where $h$ is the thickness
of the layer
in this state, so that the layer occupies the
$-h \leq \vec{m} \cdot \widetilde{\vec{x}} \leq 0$ region.

As in the classical linear case, we focus on the possible existence
of a linearly-polarized transverse wave, propagating in a direction $\vec{n}$
and polarized in a transverse direction $\vec{a}$,
both parallel to the interface. Thus, ($\vec{n}, \vec{a}, \vec{m}$) forms
an orthonormal triad.

To ensure rigid bonding the displacement must be continuous
at the interface.
The displacements $\vec{u}$ in the substrate and 
$\widetilde{\vec{u}}$ in the layer
are given respectively by
\begin{equation}
      \vec{u}=\vec{x - X}=(\vec{I} - \vec{F}^{-1})\vec{x},  \qquad
      \widetilde{\vec{u}}=\widetilde{\vec{x}} - \vec{X}=
      (\vec{I} - \widetilde{\vec{F}}^{-1})\widetilde{\vec{x}}.
      \label{eq:displ}
\end{equation}
At any point in the interface
$\vec{m \cdot x} = \vec{m} \cdot \widetilde{\vec{x}} = 0$, we have
$\vec{x}=\widetilde{\vec{x}}= \alpha \vec{n} + \beta \vec{a}$ for
some $\alpha$ and $\beta$.
Then the displacement continuity $\vec{u}= \widetilde{\vec{u}}$ at
the interface
is equivalent to $\vec{F}^{-1}(\alpha \vec{n} + \beta \vec{a})
= \widetilde{\vec{F}}^{-1}(\alpha \vec{n}  + \beta \vec{a})$.
Because this must hold for all $\alpha$, $\beta$, we conclude that
the requirement of \emph{displacement continuity} is
     \begin{equation} \label{BCdispl}
      \vec{F}^{-1}\vec{n} = \widetilde{\vec{F}}^{-1}\vec{n}, \qquad
      \vec{F}^{-1}\vec{a} = \widetilde{\vec{F}}^{-1}\vec{a}.
     \end{equation}
It then follows that
$\vec{F}^{-1}\vec{n} \times \vec{F}^{-1}\vec{a} =
\widetilde{\vec{F}}^{-1}\vec{n} \times
\widetilde{\vec{F}}^{-1}\vec{a}$. Hence, using
the identity \cite{Chad99}:
$     \vec{F}^{-1}\vec{n} \times \vec{F}^{-1}\vec{a} =
      (\det \vec{F})^{-1}\vec{F}^{T}(\vec{n}\times \vec{a})
$,
and similarly for $\widetilde{\vec{F}}$, we obtain
\begin{equation}
       (\det \vec{F})^{-1}\vec{F}^{T}\vec{m} =
        (\det \widetilde{\vec{F}})^{-1}\widetilde{\vec{F}}^{T}\vec{m},
      \label{eq:Fm}
\end{equation}
from which it follows that
\begin{equation}
      J ^{-2}\vec{m}\cdot \vec{B}\vec{m} =
      \widetilde{J}^{-2}\vec{m}\cdot \widetilde{\vec{B}}\vec{m}.
      \label{eq:mBm}
\end{equation}
We note in passing that the unit vector $\vec{M}$ normal to the faces
of the layer in the
undeformed state is given by
\begin{equation}
      \vec{M}=(\vec{m}\cdot \vec{B}\vec{m})^{-1/2}\vec{F}^{T}\vec{m} =
      (\vec{m}\cdot
\widetilde{\vec{B}}\vec{m})^{-1/2}\widetilde{\vec{F}}^{T}\vec{m},
      \label{eq:M}
\end{equation}
and that the thickness $H$ of the layer in the undeformed state is
\begin{equation}
      H= h(\vec{m}\cdot \widetilde{\vec{B}}\vec{m})^{-1/2}.
      \label{eq:H}
\end{equation}

Next, we consider the corresponding constant Cauchy stress tensors
$\vec{T}$ and $\widetilde{\vec{T}}$, in the substrate and the layer,
respectively.
The equilibrium of the  pre-strained state requires
that the upper face of the layer be subjected to the traction (deadload)
$\vec{\tau}=\widetilde{\vec{T}}\vec{m}$, and that the traction vector be
continuous at the interface, $\vec{T m} = \widetilde{\vec{T}}\vec{m}$.
Using the constitutive equations of the layer and of the substrate, this yields
\begin{align} \label{BCstress}
& \widetilde{\mu} \widetilde{J}^{-1}
       \vec{n} \cdot \widetilde{\vec{B}} \vec{m} =
       \mu J ^{-1} \vec{n \cdot B m}, \qquad
\widetilde{\mu} \widetilde{J}^{-1}
       \vec{a} \cdot \widetilde{\vec{B}} \vec{m} =
       \mu J ^{-1} \vec{a \cdot B m},   \notag \\
& \demi \widetilde{G}'(\widetilde{J})
         + \widetilde{\mu} \widetilde{J}^{-1}
             \vec{m} \cdot \widetilde{\vec{B}} \vec{m} =
       \demi G'(J) + \mu J ^{-1} \vec{m \cdot B m}.
\end{align}

Using the requirements of continuity of the displacement and of the
traction at the
interface, we now show how a given initial strain
$\widetilde{\vec{B}}$ in the layer
(resulting from a prescribed stress $\widetilde{\vec{T}}$) determines
the initial
strain $\vec{B}$ in the substrate.

First, we note that (\ref{BCstress})$_{3}$
may alternatively be written as follows, using (\ref{eq:mBm}),
\begin{equation}
      \demi G'(J) + \mu J \widetilde{J}^{-2}\vec{m} \cdot
\widetilde{\vec{B}}\vec{m}=
      \demi \widetilde{G}'(\widetilde{J}) + \widetilde{\mu}
      \widetilde{J}^{-1}\vec{m} \cdot
\widetilde{\vec{B}} \vec{m}.
      \label{eq:detJ}
\end{equation}
Because $\widetilde{\vec{B}}$ is given, $\widetilde{J}$ and
$\vec{m} \cdot \widetilde{\vec{B}} \vec{m}$ are known and so, this
is an equation for a single unknown, $J$.
It is shown in  the Appendix how the strong ellipticity
assumption implies that this
equation has at most one positive solution for $J$, and one and only
one solution if, in addition,
it is assumed that $\lim_{J\rightarrow 0}G'(J)=-\infty$.
Once $J$ is known, then (\ref{eq:mBm}) determines
$\vec{m \cdot B m}$, and (\ref{BCstress})$_{1,2}$ determine
$\vec{n \cdot B m}$ and $\vec{a \cdot B m}$. Explicitly,
\begin{multline} \label{eq:Expl1}
\vec{m \cdot B m} = (J/\widetilde{J})^{2} \vec{m} \cdot
\widetilde{\vec{B}} \vec{m},
      \qquad
\vec{n \cdot B m} =
      (J/\widetilde{J})(\widetilde{\mu}/\mu) \vec{n} \cdot
\widetilde{\vec{B}} \vec{m},
      \\
\vec{a \cdot B m} =
      (J/\widetilde{J})(\widetilde{\mu}/\mu) \vec{a} \cdot
\widetilde{\vec{B}} \vec{m}.
\end{multline}

In order to obtain all the components of $\vec{B}$ in the orthonormal triad
($\vec{n}, \vec{a}, \vec{m}$), we still need to determine $\vec{a \cdot B a}$,
$\vec{n \cdot B n}$, $\vec{n \cdot B a}$. For this purpose, we note
that the displacement
continuity requirement (\ref{BCdispl}) implies that
\begin{equation}
      \vec{n}\cdot \vec{B}^{-1}\vec{n} = \vec{n}\cdot
\widetilde{\vec{B}}^{-1}\vec{n},
      \qquad
      \vec{a}\cdot \vec{B}^{-1}\vec{a} = \vec{a}\cdot
\widetilde{\vec{B}}^{-1}\vec{a},
      \qquad
      \vec{n}\cdot \vec{B}^{-1}\vec{a} = \vec{n}\cdot
\widetilde{\vec{B}}^{-1}\vec{a}.
      \label{eq:nB-1n}
\end{equation}
Using $\vec{n}=\vec{a}\times \vec{m}$ in (\ref{eq:nB-1n})$_{1}$, and
the identity
\begin{equation}
      \vec{n}\cdot \vec{B}^{-1}\vec{n} = J ^{-2}
      [(\vec{a}\cdot \vec{B}\vec{a})(\vec{m}\cdot \vec{B}\vec{m}) -
      (\vec{a}\cdot \vec{B}\vec{m})^{2}],
      \label{eq:ID2}
\end{equation}
valid also with $\widetilde{\vec{B}}$ instead of $\vec{B}$, we obtain
\begin{equation}
      \vec{a}\cdot \vec{B}\vec{a} = \vec{a}\cdot \widetilde{\vec{B}}\vec{a} -
      \frac{(\vec{a}\cdot \widetilde{\vec{B}}\vec{m})^2}{\vec{m}\cdot
\widetilde{\vec{B}}\vec{m}} +
      \frac{(\vec{a}\cdot \vec{B}\vec{m})^2}{\vec{m}\cdot \vec{B}\vec{m}}.
      \label{eq:aBafirst}
\end{equation}
Recalling (\ref{eq:Expl1})$_{1,3}$, we obtain
\begin{equation}
      \vec{a}\cdot \vec{B}\vec{a} = \vec{a}\cdot \widetilde{\vec{B}}\vec{a} +
      \left(\frac{\widetilde{\mu}^2}{\mu ^2} - 1\right)
      \frac{(\vec{a}\cdot \widetilde{\vec{B}}\vec{m})^2}{\vec{m}\cdot
      \widetilde{\vec{B}}\vec{m}}.
      \label{eq:aBa}
\end{equation}
A similar procedure, using $\vec{a}=\vec{m}\times \vec{n}$ in
(\ref{eq:nB-1n})$_{2}$,
yields
\begin{equation}
      \vec{n}\cdot \vec{B}\vec{n} = \vec{n}\cdot \widetilde{\vec{B}}\vec{n} +
      \left(\frac{\widetilde{\mu}^2}{\mu ^2} - 1\right)
      \frac{(\vec{n}\cdot \widetilde{\vec{B}}\vec{m})^2}{\vec{m}\cdot
      \widetilde{\vec{B}}\vec{m}}.
      \label{eq:nBn}
\end{equation}
Finally, (\ref{eq:nB-1n})$_{3}$ with $\vec{n}=\vec{a}\times \vec{m}$ and
$\vec{a}=\vec{m}\times \vec{n}$ yields
\begin{equation}
      \vec{n}\cdot \vec{B}\vec{a} = \vec{n}\cdot \widetilde{\vec{B}}\vec{a} +
      \left(\frac{\widetilde{\mu}^2}{\mu ^2} - 1\right)
      \frac{(\vec{n}\cdot \widetilde{\vec{B}}\vec{m})(\vec{a}\cdot
\widetilde{\vec{B}}\vec{m})}
      {\vec{m}\cdot \widetilde{\vec{B}}\vec{m}}.
      \label{eq:nBa}
\end{equation}

To summarize: when the (constant) left Cauchy-Green strain tensor
$\widetilde{\vec{B}}$ in the layer is prescribed, then the (constant) 
left Cauchy-Green
strain tensor  $\vec{B}$ in the substrate is uniquely determined.
First, $J$ is uniquely
determined from equation (\ref{eq:detJ}). Then, all the components of $\vec{B}$
in the orthonormal triad ($\vec{n}, \vec{m}, \vec{a}$) are
explicitly given by equations
(\ref{eq:Expl1})$_{1,2,3}$, (\ref{eq:aBa}), (\ref{eq:nBn}), (\ref{eq:nBa}).

In order to use the exact wave solutions described in Section \ref{section3},
we shall assume from now on that the condition (\ref{nBm}) is
fulfilled in the layer, which, by
(\ref{eq:Expl1})$_2$ implies that it is also fulfilled in the substrate~:
\begin{equation} \label{nBmTilde}
      \vec{n} \cdot\vec{B}\vec{m}=0,\qquad
      \vec{n} \cdot\widetilde{\vec{B}}\vec{m} = 0.
\end{equation}
Then, the equations for the other components of $\vec{B}$ reduce to
\begin{align}  \label{Expl2}
&     \vec{m \cdot B m} = (J/\widetilde{J})^{2} \vec{m} \cdot
\widetilde{\vec{B}} \vec{m},
      &&
      \vec{a \cdot B m} =
      (J/\widetilde{J})(\widetilde{\mu}/\mu) \vec{a} \cdot
\widetilde{\vec{B}} \vec{m},
      \nonumber \\
  &    \vec{a}\cdot \vec{B}\vec{a} = \vec{a}\cdot \widetilde{\vec{B}}\vec{a} +
      \left(\frac{\widetilde{\mu}^2}{\mu ^2} - 1\right)
      \frac{(\vec{a}\cdot \widetilde{\vec{B}}\vec{m})^2}{\vec{m}\cdot
      \widetilde{\vec{B}}\vec{m}}, &&
      \vec{n}\cdot \vec{B}\vec{n} = \vec{n}\cdot
\widetilde{\vec{B}}\vec{n}, \notag \\
 &     \vec{n}\cdot \vec{B}\vec{a} = \vec{n}\cdot
\widetilde{\vec{B}}\vec{a}.
&& \end{align}
In tensorial form, the expression of
$\vec{B}$ in terms of $\widetilde{\vec{B}}$ is given by
\begin{multline} \label{ExplTens}
(\vec{m}\cdot \widetilde{\vec{B}}\vec{m})\vec{B} =
(\vec{m}\cdot \widetilde{\vec{B}}\vec{m})
\widetilde{\vec{B}}-\widetilde{\vec{B}}\vec{m}\otimes
\widetilde{\vec{B}}\vec{m} \\
  +  \left[(\widetilde{\mu}/\mu)(\vec{a}\cdot
\widetilde{\vec{B}}\vec{m})\vec{a}+
(J/\widetilde{J})(\vec{m}\cdot
\widetilde{\vec{B}}\vec{m})\vec{m}\right] \\ \otimes
\left[(\widetilde{\mu}/\mu)(\vec{a}\cdot \widetilde{\vec{B}}\vec{m})\vec{a}+
(J/\widetilde{J})(\vec{m}\cdot
\widetilde{\vec{B}}\vec{m})\vec{m}\right].
\end{multline}
In particular, we note that $\vec{Bn}=\widetilde{\vec{B}}\vec{n}$.

%+++++++++++++++++++++++++++++++++++++

\section{Large amplitude Love wave}
\label{section5}

%+++++++++++++++++++++++++++++++++++++

Now we look at wave propagation in the initially deformed structure.
In the substrate, we require that amplitude of the wave decays in the direction
of $\vec{m}$, and hence the displacement field $\vec{u}$ is assumed
to be of the
form (\ref{Att}). In the layer, we consider an unattenuated 
time-harmonic displacement field
$\widetilde{\vec{u}}$ of the form (\ref{Unatt}). As in the classical case,
we wish to combine these exact wave solutions in order to obtain a global
time-harmonic wave motion with propagation speed $v$. Note that both
displacement fields
need to be of the same angular frequency, and hence the same wavenumber,
in order to satisfy
boundary conditions at the interface. Thus, using
(\ref{Att}) for the substrate and (\ref{Unatt}) for the layer, we write
\begin{equation} \label{solSubs}
\vec{u}(\vec{x}, t) = A \exp \left(-\frac{\gamma}{v_{ 
\vec{m}}}\vec{m}\cdot \vec{x}\right)
\cos k(\vec{n} \cdot \vec{x} - v t) \vec{a},
\end{equation}
and
\begin{equation} \label{solLayer}
\widetilde{\vec{u}}(\vec{x}, t) =
     \left[B\sin(\frac{\kappa}{\widetilde{v}_{ \vec{m}}}
\vec{m} \cdot \vec{x})+C\cos(\frac{\kappa}{\widetilde{v}_{\vec{m}}}
\vec{m} \cdot \vec{x})\right]
        \cos k(\vec{n} \cdot \vec{x} -v t) \vec{a},
        \end{equation}
where
\begin{equation}
      k= \frac{\kappa}{\sqrt{v^2-\widetilde{v}^2_{\vec{n}}}}
      =\frac{\gamma}{\sqrt{v^{2}_{\vec{n}}-v^2}},
      \label{eq:k}
\end{equation}
is the \emph{wavenumber}.
Here, in accordance with (\ref{homogv}), the body waves
speeds $v_{\vec n}$,
$v_{\vec m}$, $\widetilde{v}_{\vec n}$, $\widetilde{v}_{\vec m}$ are given by
\begin{multline}
     \rho_0 v^2_{\vec m} = \mu \, \vec{m \cdot B m}, \qquad
     \rho_0 v^2_{\vec n} = \mu \, \vec{n \cdot B n}, \\
     \widetilde{\rho}_0 \widetilde{v}^2_{\vec m} =
     \widetilde{\mu} \, \vec{m} \cdot \widetilde{\vec{B}} \vec{m}, \qquad
     \widetilde{\rho}_0 \widetilde{v}^2_{\vec n} =
     \widetilde{\mu} \, \vec{n} \cdot \widetilde{\vec{B}} \vec{n},
      \label{body}
\end{multline}
and the Love wave speed $v$ has to satisfy
\begin{equation} \label{eq: ineq}
\widetilde{v}^2_{\vec n} < v^2 < v^2_{\vec n}.
\end{equation}
Notice that this is possible only when $v^2_{\vec n} >
\widetilde{v}^2_{\vec n}$, or
equivalently, recalling $\vec{n \cdot B n}=\vec{n} \cdot
\widetilde{\vec{B}} \vec{n}$, when
$\mu/\rho_0 > \widetilde{\mu}/\widetilde{\rho}_0$. Thus, Love waves require
the combination of a `slow' (or `soft') layer over a
`fast' (or `hard') substrate, independently of the initial pre-strain.

We now show that the boundary conditions may be satisfied. This leads to
the dispersion equation (a relation between $k$ and $v$) and the determination
of the constants $A$, $B$, $C$ in terms of a single parameter characterizing
the amplitude of the wave.

The first boundary condition to enforce is that the displacement is
continuous at the layer/substrate interface $\vec{m \cdot x} = 0$.
Using \eqref{solSubs} and \eqref{solLayer}, this gives
\begin{equation} \label{BCa}
A = C.
\end{equation}

The second boundary condition is the continuity of the traction vector
at the layer/substrate interface $\vec{m}\cdot \vec{x}=0$. Using
(\ref{eq:traction}), and
applying it to the wave motion \eqref{solSubs}, we obtain, for the
traction $\vec{t}$
on a plane $\vec{m}\cdot \vec{x}= \mbox{\it constant}\,$ in the substrate,
\begin{equation} \label{eq:tSubs}
     \vec{t} = \vec{T m} - \gamma  \rho v_{\vec m}  A
      \exp \left(-\frac{\gamma}{v_{ \vec{m}}}\vec{m}\cdot \vec{x}\right)
       \cos k(\vec{n} \cdot \vec{x} -v t) \vec{a},
\end{equation}
where  $\rho = J^{-1} \rho_0$
is the mass density of the substrate in the intermediate configuration.
Similarly for the traction $\widetilde{\vec{t}}$
on a plane $\vec{m}\cdot \vec{x}= \mbox{\it constant}\,$ in the layer, we find
\begin{equation} \label{eq:tLayer}
\widetilde{\vec{t}} =
      \widetilde{\vec{T}} \vec{m}
      +  \kappa \widetilde{\rho} \, \widetilde{v}_{\vec m}
         \left[B\cos\left(\frac{\kappa}{\widetilde{v}_{\vec{m}}} 
\vec{m} \cdot \vec{x}\right)-
     C\sin\left(\frac{\kappa}{\widetilde{v}_{ \vec{m}}}
\vec{m} \cdot \vec{x}\right)\right]
        \cos k\left(\vec{n} \cdot \vec{x} -v t\right) \vec{a}.
\end{equation}
where  $\widetilde{\rho} = J^{-1} \widetilde{\rho}_0$
is the mass density of the layer in the intermediate configuration.
Also, recall that $\vec{T m}=\widetilde{\vec{T}} \vec{m}=\vec{\tau}$, where
$\vec{\tau}$ is the constant traction (deadload) applied at the
upper face of the layer. Hence the condition $\vec{t}=\widetilde{\vec{t}}$
at the layer/substrate interface $\vec{m}\cdot \vec{x}=0$ reads
\begin{equation} \label{BCb}
\gamma \rho v_{\vec m} A +\kappa \widetilde{\rho} \,
\widetilde{v}_{\vec m} B = 0.
\end{equation}

The third boundary condition is that, at the upper face of the
layer $\vec{m}\cdot \vec{x}=-h$, the wave creates no traction in addition to
the static traction (deadload) $\vec{\tau}$. Using (\ref{eq:tLayer}),
this yields
\begin{equation} \label{BCc}
B\cos(\frac{\kappa}{\widetilde{v}_{\vec{m}}} h) +
     C\sin(\frac{\kappa}{\widetilde{v}_{ \vec{m}}} h) = 0.
\end{equation}

Equations \eqref{BCa}, \eqref{BCb}, \eqref{BCc} form an algebraic
linear homogeneous
system for the three unknowns $A$, $B$, $C$. Writing the condition
for non-trivial solutions
and using \eqref{eq:k}, we arrive at the following \emph{dispersion
equation} relating the
wave speed $v$ to the wave number $k$,
\begin{equation} \label{disp}
\tan \left[ k h \sqrt{\frac{v^2 -
\widetilde{v}^2_{\vec n}}{\widetilde{v}^2_{\vec{m}}}} \right]
-\frac{\rho v_{\vec{m}}}{\widetilde{\rho} \, \widetilde{v}_{\vec{m}}}
        \sqrt{\frac{v^2_{\vec n} - v^2}{v^2 - \widetilde{v}^2_{\vec{n}}}} = 0.
\end{equation}

Let $c$ and $\widetilde{c}$ denote the transverse bulk wave speeds in
the underformed substrate and layer, respectively:
$c^2=\mu / \rho_0$,  $\widetilde{c}^2=\widetilde{\mu} / \widetilde{\rho}_0$.
Using \eqref{Expl2} and \eqref{body}, we note that
$v^2_{\vec n}/\widetilde{v}^2_{\vec n}=c^2/\widetilde{c}^2$ and
$\rho v_{\vec m}/(\widetilde{\rho} \widetilde{v}_{\vec m})=
\rho_0 c/(\widetilde{\rho}_0 \widetilde{c})$, so that the dispersion
equation \eqref{disp}
may also be written as
\begin{equation} \label{dispAlt}
\tan \left[ k h (\widetilde{v}_{\vec n}/\widetilde{v}_{\vec m})
\sqrt{(v/\widetilde{v}_{\vec n})^2 - 1} \right]
-\frac{\rho_0 c}{\widetilde{\rho}_0 \widetilde{c}}
        \sqrt{\frac{(c/\widetilde{c})^2 - (v/\widetilde{v}_{\vec n})^2}
        {(v/\widetilde{v}_{\vec n})^2 - 1}} = 0.
\end{equation}

\begin{figure}
\centering
\epsfig{figure=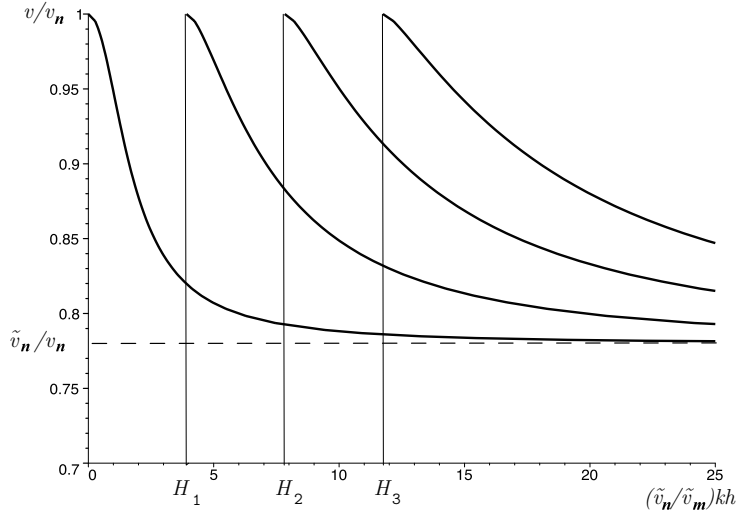, width=.7\textwidth}
   \caption{
   Dispersion curves for successive Love wave modes in a pre-strained 
configuration.
   Here $H_l=l\pi/\sqrt{(c/\widetilde{c})^2-1}$, ($l=1,2, 3,4, 
\ldots$). Thus, for
   $0< kh <(\widetilde{v}_{\vec m}/\widetilde{v}_{\vec n})H_1$,
   only one mode may propagate (fundamental mode). For
   $(\widetilde{v}_{\vec m}/\widetilde{v}_{\vec n})H_l< kh
   <(\widetilde{v}_{\vec m}/\widetilde{v}_{\vec n})H_{l+1}$,
   $l+1$ modes may propagate.}
   \label{dispersioncurves}
\end{figure}

In the absence of pre-strain,  $\vec{B} =\widetilde{\vec{B}}=
\vec{I}$, $J = \widetilde{J} = 1$,
hence $v^2_{\vec n} = v^2_{\vec m} = c^2$ in the substrate and
$\widetilde{v}^2_{\vec n} = \widetilde{v}^2_{\vec m} =
\widetilde{c}^2$ in the layer,
so that this dispersion equation specializes to
\begin{equation} \label{disp_iso}
\tan \left[ k h \sqrt{(v/\widetilde{c})^2 - 1} \right]
-\frac{\rho_0 c}{\widetilde{\rho}_0 \widetilde{c}}
        \sqrt{\frac{(c/\widetilde{c})^2 - (v/\widetilde{c})^2}
        {(v/\widetilde{c})^2 - 1}} = 0,
\end{equation}
which coincides with the dispersion equation of linear isotropic 
elasticity \color{black}
\cite {EJP57}. \color{black}From a practical point of view, any 
dispersion curve obtained from
\eqref{disp_iso}
as a plot of $v/\widetilde{c}$ against $k h$ for a given choice of
$\rho_0 / \widetilde{\rho}_0$ and $c / \widetilde{c}$ can be used in the
present context of \eqref{dispAlt},
by identifying $v/\widetilde{c}$ with $v/\widetilde{v}_{\vec n}$ and
$k h$ with $(\widetilde{v}_{\vec n} / \widetilde{v}_{\vec m}) k h$,
see Willson \cite{Will75} for similar results in the small-on-large theory.
A typical plot for the different wave modes is presented in 
Fig.\ref{dispersioncurves}.
Of course, the scope of the results is now richer because they include
large amplitudes and pre-stress.

When the dispersion equation \eqref{disp} is satisfied,
the solution of the linear homogeneous system \eqref{BCa},
\eqref{BCb}, \eqref{BCc}
for $A$, $B$, $C$ is
\begin{equation} \label{SolABD}
      A=\alpha \cos\left(\frac{\kappa}{\widetilde v_{\vec{m}}} h\right), \qquad
      B=- \alpha \sin\left(\frac{\kappa}{\widetilde v_{\vec{m}}} 
h\right), \qquad
      C=\alpha\cos\left(\frac{\kappa}{\widetilde v_{\vec{m}}} h\right),
\end{equation}
where $\alpha$ is arbitrary. Thus $A$, $B$, $C$ are expressed in
terms of a single
parameter $\alpha$ characterizing the amplitude of the Love wave.

Finally, we now denote by $(\eta, \xi, \zeta)$ the Cartesian coordinates
along ($\vec{n}, \vec{a}, \vec{m}$),
\begin{equation}
\eta =  \vec{n} \cdot \vec{x}, \qquad \xi = \vec{a} \cdot \vec{x} , \qquad
\zeta = \vec{m}\cdot \vec{x},
\end{equation}
and we find that the displacement
fields (\ref{solSubs}) (\ref{solLayer})
are
\begin{align}
& \vec{u} = \alpha \cos\left(\frac{\kappa}{\widetilde v_{\vec{m}}} h\right)
\exp \left(-\frac{\gamma}{v_{ \vec{m}}}\zeta\right)
\cos k(\eta - v t) \vec{a} , \notag \\
& \widetilde{\vec{u}}  =  \alpha \cos\frac{\kappa}{\widetilde
v_{\vec{m}}} (h+ \zeta)
\cos k( \eta -v t) \vec{a}, \label{Newsol}
\end{align}
or equivalently, recalling \eqref{eq:k},
\setlength{\arraycolsep}{2pt}
\begin{eqnarray}
\vec{u} & = & \alpha \cos \left[ k h \sqrt{\frac{v^2 -
\widetilde{v}^2_{\vec n}}{\widetilde{v}^2_{\vec{m}}}} \right]
\exp \left[- k \zeta \sqrt{\frac{v^2_{\vec n} -
v^2}{v^2_{\vec{m}}}} \right] \cos k(\eta - v t) \vec{a}
\label{NewsolSubsAlt}  , \\
\widetilde{\vec{u}} & = & \alpha \cos \left[ k (h + \zeta) \sqrt{\frac{v^2 -
\widetilde{v}^2_{\vec n}}{\widetilde{v}^2_{\vec{m}}}} \right]
\cos k( \eta -v t) \vec{a}. \label{NewsolLayerAlt}
\end{eqnarray}

%-------------------------------------------

\section{Energy density and energy flux}
\label{section6}

%-------------------------------------------

Here we compute  the energy flux and the energy density associated with
a motion of the type (\ref{eq:displace}) in compressible neo-Hookean materials.
Let $W$ and $\overline{W}$ be the strain energy densities corresponding,
respectively, to the static deformation (\ref{eq:3.1}) and to the
motion (\ref{eq:displace}), both measured per unit volume of the
undeformed state. Then,
the \emph{energy density} ${\cal E}$, measured per unit volume of the 
homogeneously
deformed state (\ref{eq:3.1}), and the corresponding \emph{energy flux vector}
$\bR$ are \cite{BHT94}
\begin{equation}
{\cal E} = \demi \rho\dot{\overline{\vec{x}}}
\cdot\dot{\overline{\vec{x}}}
+ J^{-1}(\overline{W}-W), \label{eq:6.1} \qquad
{\cal R}_k = -\dot{\overline{x}}_i\overline{P}_{ik},
\end{equation}
where the Piola-Kirchhoff stress tensor $\overline{\vec{P}}$ with respect to
the state of homogeneous static deformation is defined by (\ref{eq:3.16}).
They satisfy the energy balance equation
\begin{equation}
(\partial {\cal E}/\partial t)+(\partial {\cal R}_k/\partial x_k)=0,
\label{eq:6.3}
\end{equation}
where $x_k$ are the coordinates in the state of homogeneous static deformation,
and the partial derivative with respect to time is taken at
fixed $\vec{x}$.

We now evaluate the energy density and the energy flux vector for the
wave motion \eqref{Newsol}$_2$ in the layer, and for the wave motion 
\eqref{Newsol}$_1$
in the substrate.
Using \eqref{Newsol}$_2$, we obtain for the layer
\begin{eqnarray}
\widetilde{{\cal E}} &=& \textstyle{\frac{1}{2}} \widetilde \rho 
k^2\alpha^2 (v^2+\widetilde
v^2_{\vec{n}})\cos^2 \frac{\kappa}{\widetilde v_{
\vec{m}}}(h+\zeta)\sin^2 k(\eta-vt)  \nonumber \\
           && + \textstyle{\frac{1}{2}} \widetilde \rho k^2\alpha^2 
(v^2-\widetilde
v^2_{\vec{n}})\sin^2 \frac{\kappa}{\widetilde v_{
\vec{m}}}(h+\zeta)\cos^2 k(\eta-vt)  \label{EL} \\
           && - \alpha\ \vec{a}\!\cdot\! [\frac{\kappa}{\widetilde v_{\vec{m}}}
\sin \frac{\kappa}{\widetilde v_{\vec{m}}}(h+\zeta)\cos
k(\eta-vt)\widetilde{\vec{T}}\vec{m} \notag \\ &&
+ k \cos \frac{\kappa}{\widetilde v_{\vec{m}}}(h+\zeta)\sin k(\eta-vt)\widetilde{\vec{T}}\vec{n}], \nonumber \\
\widetilde{\bR}&=&\cos\frac{\kappa}{\widetilde v_{\vec{m}}}(h+\zeta)\sin
k(\eta-vt)\{
-vk\alpha\widetilde{\vec{T}}\vec{a}\label{FL} \\
        && + \alpha^2 kv \widetilde{\mu} \widetilde{J}^{-1}
[\frac{\kappa}{\widetilde v_{\vec{m}}}
\sin \frac{\kappa}{\widetilde v_{\vec{m}}}(h+\zeta)\cos k(\eta-vt)
\widetilde{\vec{B}}\vec{m} \notag \\
&&        + k\cos\frac{\kappa}{\widetilde v_{\vec{m}}}(h+\zeta)\sin k(\eta-vt)
\widetilde{\vec{B}} \vec{n}]\}. \nonumber
\end{eqnarray}

Using \eqref{Newsol}$_1$, we obtain for the substrate
\begin{eqnarray}
{\cal E} &=& \textstyle{\frac{1}{2}}\rho k^2\alpha^2\cos^2 
(\frac{\kappa}{\widetilde
v_{\vec{m}}}h)\mbox{exp}(\frac{-2\gamma}{v_{\vec{m}}}\zeta)[v^2_{\vec{n}} +
v^2\sin^2k(\eta-vt)-v^2\cos^2k(\eta-vt)] \nonumber \\
           && - \alpha\cos(\frac{\kappa}{\widetilde v_{\vec{m}}}
h)\mbox{exp}(\frac{-\gamma}{v_{\vec{m}}}\zeta)\ \vec{a}\!\cdot\!
[\frac{\gamma}{v_{
\vec{m}}} \cos k(\eta-vt)\vec{T}\vec{m} \notag \\
&&           + k\sin k(\eta-vt)\vec{T}\vec{n}], \; \label{ES} \\
       \bR &=& -vk\alpha \cos(\frac{\kappa}{\widetilde v_{\vec{m}}}h)\sin
k(\eta-vt)\mbox{exp}(\frac{-\gamma}{v_{ \vec{m}}}\zeta)\vec{T}\vec{a}
\label{FS} \\
           && + \alpha^2 kv \mu J^{-1} \cos^2 (\frac{\kappa}{\widetilde
v_{\vec{m}}}h)\mbox{exp}(\frac{-2\gamma}{v_{\vec{m}}}\zeta)[\frac{\gamma}{v_{
\vec{m}}}\cos k(\eta -vt)\vec{B}\vec{m} \notag \\
&& + k \sin
k(\eta-vt)\vec{B}\vec{n}]. \nonumber
\end{eqnarray}

Mean energy densities and mean energy fluxes are obtained by 
averaging over a period
in time at fixed $\vec{x}$,
\begin{equation}
\langle {\cal E} \rangle =(\omega /2\pi)\int_{0}^{2\pi/\omega}
{\cal E}(\vec{x}, t)\, dt , \label{ME} \qquad
\langle {\bR} \rangle = (\omega /2\pi)\int_{0}^{2\pi/\omega}
\bR(\vec{x}, t)\, dt .
\end{equation}

Using (\ref{EL}), (\ref{FL}), (\ref{ME}),  we find the mean energy
density and the mean energy flux in the layer as
\begin{align}
& \langle \widetilde{{\cal E}}\rangle =  \textstyle{\frac{1}{4}} 
\widetilde\rho k^2 \alpha ^2
\{v^2+\widetilde v^2_{\vec{n}} [\cos^2 \frac{\kappa}{\widetilde
v_{\vec{m}}}(h+ \zeta)
-\sin ^2 \frac{\kappa}{\widetilde v_{\vec{m}}} (h+\zeta)]\}, \notag \\
& \langle \widetilde{{\bR}} \rangle =  \textstyle{\frac{1}{2}}vk^2\alpha^2
\widetilde{\mu} \widetilde{J}^{-1}
\cos^2 \frac{\kappa}{\widetilde v_{\vec{m}}} (h+\zeta)
\ \widetilde{\vec{B}} \vec{n}. \label{MFL}
\end{align}

Using (\ref{ES}), (\ref{FS}), (\ref{ME}), we find the mean energy
density and the mean energy flux in the substrate as
\begin{align}
& \langle {\cal E} \rangle =  \textstyle{\frac{1}{2}}\rho v^2_{\vec{n}}
k^2\alpha^2\cos^2(\frac{\kappa}{\widetilde v_{\vec{m}}}
h)\mbox{exp}(\frac{-2\gamma}{v_{\vec{m}}} \zeta), \notag \\
& \langle {\bR} \rangle =  \textstyle{\frac{1}{2}}v k^2 \alpha^2
\mu J^{-1} \cos^2(\frac{\kappa}{\widetilde v_{\vec{m}}} h) \
\mbox{exp}(\frac{-2\gamma}{v_{\vec{m}}} \zeta)\ \vec{B} \vec{n}. \label{MFS}
\end{align}
The mean energy flux in the layer and the mean energy flux in the
substrate are both
along $\widetilde{\vec{B}}\vec{n}=\vec{Bn}$, thus along the same
direction, parallel to the interface.
In general, this direction is not along the propagation
direction $\vec{n}$.
This is an effect of the anisotropy induced by the pre-strain.
However, in the special
case when $\vec{n}$ is along a principal direction of $\vec{B}$ and
$\widetilde{\vec B}$, the mean energy fluxes are along $\vec{n}$.

Also, we note that at the interface $\zeta=0$, the mean energy flux
$\langle {\bR} \rangle _0$ (say) in the substrate is related to the
mean energy flux $\langle \widetilde{{\bR}} \rangle _0$ (say) in the 
layer through
\begin{equation}
\widetilde{\mu}^{-1}\widetilde{J} \langle \widetilde{{\bR}} \rangle _0 =
\mu^{-1}J \langle {\bR} \rangle _0.
      \label{Finter}
\end{equation}
Similarly, for the mean energy densities $\langle {\cal E} \rangle _0$ and
$\langle \widetilde{{\cal E}}\rangle _0$ at the interface $\zeta=0$ , we find
\begin{equation}
      \widetilde{\mu}^{-1}\widetilde{J}
      \frac{\langle \widetilde{{\cal E}}\rangle _0}
      {(v/\widetilde{v}_{\vec n})^2 +
      \cos (2\frac{\kappa}{\widetilde v_{\vec{m}}} h)} =
      \mu^{-1}J \frac{\langle {\cal E} \rangle _0}
      {1+\cos (2\frac{\kappa}{\widetilde v_{\vec{m}}} h)}.
      \label{Einter}
\end{equation}
Recalling that $v^2 >\widetilde{v}^2_{\vec n}$, we note in particular that
$\widetilde{\mu}^{-1}\widetilde{J}\langle \widetilde{{\cal E}}\rangle _0 >
\mu^{-1}J \langle {\cal E} \rangle _0$.

We now consider the \emph{energy flux velocity} defined as the mean energy
flux vector divided by
the mean energy density. For the energy flux velocity
$\widetilde{\vec g}$ wave in the layer,
we have
\begin{equation}
\widetilde{\vec g} = \frac{\langle \widetilde{{\bR}} \rangle}
{\langle \widetilde{{\cal E}}\rangle} =
v\, \frac{1+\cos 2\frac{\kappa}{\widetilde v_{\vec{m}}}(h+\zeta)}
{(v/\widetilde{v}_{\vec n})^2 + \cos 2\frac{\kappa}{\widetilde
v_{\vec{m}}}(h +\zeta)}
\, \frac{\widetilde{\vec{B}} \vec{n}}{(\vec{n} \cdot
\widetilde{\vec{B}} \vec{n})},
\label{gtilde}
\end{equation}
and for the  energy flux velocity $\vec{g}$ in the substrate, we have
\begin{equation}
\vec{g} =\frac{\langle {\bR}\rangle}{\langle {\cal E} \rangle} =
v\,  \frac{\vec{B} \vec{n}}{(\vec{n} \cdot \vec{B} \vec{n})}.
\label{g}
\end{equation}
In the layer, the energy flux velocity depends on the depth $\zeta$
whilst in the substrate,
the energy flux velocity is the same at all points.
This is because the wave motion in the substrate consists of a 
\emph{single train} of
inhomogeneous plane
waves. On the contrary, the wave motion in the layer may be viewed as
a \emph{superposition of trains} of homogeneous plane waves.
Because $\widetilde{\vec{B}}\vec{n}=\vec{Bn}$, the energy flux
velocities are related through
\begin{equation}
[(v/\widetilde{v}_{\vec n})^2 + \cos 2\frac{\kappa}{\widetilde
v_{\vec{m}}}(h +\zeta)]
\widetilde{\vec g} = [1+\cos 2\frac{\kappa}{\widetilde
v_{\vec{m}}}(h+\zeta)]\vec{g}.
\label{gg}
\end{equation}
Also, because $v^2 >\widetilde{v}^2_{\vec n}$, we note that
the energy flux velocity at any point of the layer is
smaller in magnitude that the energy flux velocity in the substrate.
Finally we note that
\begin{equation}
\vec{g}\cdot \vec{n} = v ,  \qquad \vec{g}\cdot \vec{m} = 0 ,
\label{Hproperties}
\end{equation}
in accordance with previous results about finite amplitude
inhomogeneous plane waves in
unbounded deformed Blatz-Ko materials \cite{1} \cite{2}. These
relations are the
same as those derived by Hayes in the context of linear theories
\cite{HEnergy}.

We now define \emph{total mean energy densities} and \emph{total mean
energy fluxes} as
\begin{align}
&  \langle \widetilde{{\cal E}}\rangle _{T}  = \int^{0}_{-h}\langle
\widetilde{{\cal E}}(\zeta) \rangle \text{d}\zeta,
\  \langle \widetilde{{\bR}}\rangle _{T} = \int^{0}_{-h}\langle
\widetilde{{\bR}}(\zeta)\rangle  \text{d}\zeta , \notag \\
&  \langle {\cal E}\rangle _{T}   = \int^{\infty}_{0}\langle {\cal 
E}(\zeta) \rangle  \text{d}\zeta,
\  \langle {\bR}\rangle _{T} = \int^{\infty}_{0}\langle 
{\bR}(\zeta)\rangle  \text{d}\zeta .
\end{align}
The total mean energy density $\langle \widetilde{{\cal
E}}\rangle _{T}$ is the wave energy
in the layer $(-h < \zeta =\vec{m}\cdot \vec{x} <0)$ per unit length
(along $\vec{n}$)
and per unit width (along $\vec{a}$) of the layer. Similarly, the
total mean energy density
$\langle {\cal E}\rangle _{T}$ is the energy  in the substrate
$(0 < \zeta =\vec{m}\cdot \vec{x} < \infty)$ per unit length (along
$\vec{n}$) and per unit width
(along $\vec{a}$) of the substrate.
The total mean energy flux $\langle \widetilde{{\bR}}\rangle _{T}$
is the energy flux characterizing the rate at which energy flows
through a normal section of the layer
$(-h < \zeta =\vec{m}\cdot \vec{x} <0)$ per unit width of this section.
Similarly, the total mean energy flux $\langle {\bR}\rangle _{T}$
is the energy flux characterizing the rate at which energy flows
through a normal section of the
substrate $(-h < \zeta =\vec{m}\cdot \vec{x} <0)$ per unit width of
this section.

For the wave motion in the layer, we obtain
\begin{align}
& \langle \widetilde{{\cal E}}\rangle _{T}  =
\textstyle{\frac{1}{4}}\widetilde{\rho}k^2\alpha^2\left[\widetilde{v}^{2}_{\vec{n}}
\frac{\widetilde 
v_{\vec{m}}}{2\kappa}\sin(2\frac{\kappa}{\widetilde{v}_{\vec{m}}}h)+v^2h\right],
\notag \\
& \langle \widetilde{{\bR}}\rangle _{T}  =
\textstyle{\frac{1}{4}}v 
k^2\alpha^2\left[\frac{\widetilde{v}_{\textbf{m}}}{2\kappa}\sin(2\frac{\kappa}
{\widetilde{v}_{\textbf{m}}}h)+h\right]\widetilde{\mu}
\widetilde{J}^{-1}\widetilde{\textbf{B}}\textbf{n},
\end{align}
and for the wave motion in the substrate, we obtain
\begin{equation}
\langle {\cal E}\rangle _{T}  = \textstyle{\frac{1}{4}}\rho
k^2\alpha^2 v^{2}_{\vec{n}}\frac{v_{\vec{m}}}{\gamma}
\cos^2(\frac{\kappa}{\widetilde{v}_{\vec{m}}}h), \qquad
\langle {\bR}\rangle _{T} =
\textstyle{\frac{1}{4}}k^2\alpha^2v\frac{v_{\vec{m}}}{\gamma}
\cos^2(\frac{\kappa}{\widetilde{v}_{\textbf{m}}}h) \mu J^{-1}\vec{B}\vec{n}.
\end{equation}
Here we note that the repartition of energy between the layer and the
substrate depends on
the depth $h$ of the layer, or more precisely, on the dimensionless
parameter $k h$ characterizing the ratio of the layer depth to the wavelength.

%***************************************

\section{Interface in a principal plane}
\label{section7}

%***************************************

For a given static strain $\widetilde{\vec B}$ in the layer and a
given unit vector
$\vec{m}$, there is, in general, only one direction $\vec{n}$ in the interface
$\vec{m}\cdot \vec{x}=\vec{m}\cdot \widetilde{\vec x}=0$ along which
a finite amplitude
Love wave as described in Section \ref{section5} may propagate. Indeed, because
$\vec{n}\cdot \vec{m}=\vec{n}\cdot \widetilde{\vec B}\vec{m}=0$ is required,
$\vec{n}$ must be along $\vec{m}\times \widetilde{\vec B}\vec{m}$.
However, if $\vec{m}$ is along a principal axis of $\widetilde{\vec B}$,
then $\vec{n}\cdot \vec{m}=\vec{n}\cdot \widetilde{\vec B}\vec{m}=0$ is
satisfied automatically for any propagation direction
$\vec{n}$ orthogonal to $\vec{m}$ that is, $\vec{n}$ can be along any 
direction in the interface.
Here, we consider this special case
and give a numerical
example showing the effects of strain-induced anisotropy on the wave
characteristics.

Calling $\vec{i}, \vec{j}, \vec{k}$ the unit vectors along the principal
axes of $\widetilde{\vec B}$, we write
\begin{equation} \label{namtheta}
\vec{n} = \cos \theta \vec{i} + \sin \theta \vec{j}, \qquad
\vec{a} = -\sin \theta \vec{i} + \cos \theta \vec{j}, \qquad
\vec{m} = \vec{k},
\end{equation}
where the angle $\theta \in [0, 2\pi]$ is arbitrary. The left
Cauchy-Green strain tensor
in the layer is
\begin{equation}
      \widetilde{\vec B}=\widetilde{\lambda}_1^2 \vec{i}\otimes \vec{i} +
      \widetilde{\lambda}_2^2 \vec{j}\otimes \vec{j} +
      \widetilde{\lambda}_3^2 \vec{k}\otimes \vec{k},
      \label{BtildeDiag}
\end{equation}
where $\widetilde{\lambda}_1$,  $\widetilde{\lambda}_2$,
$\widetilde{\lambda}_3$
are the principal stretches in the layer. The left Cauchy-Green strain tensor
$\vec{B}$ in the substrate is then uniquely determined as explained in Section
\ref{section4}. First $J$ is determined from equation
\eqref{eq:detJ}, which here reads
\begin{equation}
      \demi G'(J) + \mu
\widetilde{\lambda}^{-2}_{1}\widetilde{\lambda}^{-2}_2 J =
      \demi \widetilde{G}'(\widetilde{J}) + \widetilde{\mu}
\widetilde{\lambda}^{-1}_{1}\widetilde{\lambda}^{-1}_2\widetilde{\lambda}_3,
      \label{detJlambda}
\end{equation}
with
$\widetilde{J}=\widetilde{\lambda}_1\widetilde{\lambda}_2\widetilde{\lambda}_3$.
Then, using \eqref{ExplTens}, we obtain
\begin{equation}
      \vec B=\widetilde{\lambda}_1^2 \vec{i}\otimes \vec{i} +
      \widetilde{\lambda}_2^2 \vec{j}\otimes \vec{j} +
      \lambda_3^2 \vec{k}\otimes \vec{k},
      \label{BDiag}
\end{equation}
where $\lambda_{3}$ is given by
\begin{equation}
      \lambda_3^{2} =
\widetilde{\lambda}^{-2}_1\widetilde{\lambda}^{-2}_2 J^2.
      \label{lambda3}
\end{equation}

Hence, the \emph{dispersion equation} relating the
wave speed $v$ and the wave number $k$ is \eqref{disp}, or,
equivalently, \eqref{dispAlt},
where
\begin{equation}
      \widetilde{v}^2_{\vec n}/\widetilde{c}^2 = v^2_{\vec n}/c^2 =
      \widetilde{\lambda}^{2}_1 \cos^2 \theta +
\widetilde{\lambda}^{2}_2 \sin^2 \theta,
      \qquad \widetilde{v}^2_{\vec 
m}/(\widetilde{c}^2\widetilde{\lambda}^{2}_3) =
      v^2_{\vec m}/(c^2\lambda^{2}_3) = 1 .
      \label{vtheta}
\end{equation}

We note that when both the substrate and the layer are of
the Levinson  and  Burgess type \eqref{LevBurg} with Lam\'e
parameters $\lambda$, $\mu$,
and $\widetilde{\lambda}$, $\widetilde{\mu}$, respectively, the equation
\eqref{detJlambda} for the determination of $J$ reduces to the linear equation
\begin{equation}
      (\lambda + \mu + \mu \widetilde{\lambda}^{-2}_1
\widetilde{\lambda}^{-2}_2)J
      - (\lambda + 2\mu) =
       (\widetilde{\lambda} + \widetilde{\mu} + \widetilde{\mu}
       \widetilde{\lambda}^{-2}_1 \widetilde{\lambda}^{-2}_2)
       \widetilde{\lambda}_1\widetilde{\lambda}_2\widetilde{\lambda}_3
      - (\widetilde{\lambda} + 2\widetilde{\mu}).
      \label{LinJ}
\end{equation}

We now present a numerical example. We take both the layer and the substrate
to be of the Levinson  and  Burgess type \eqref{W}-\eqref{LevBurg}, with
   $\lambda = \mu$ and $\widetilde{\lambda} = \widetilde{\mu}$,
an assumption often encountered in the geophysics literature
(it leads to an infinitesimal Poisson ratio of $1/4$,
which is common for rocks).
Hence, $G'(J) = 2 \mu (2J - 3)$ and $\widetilde{G}'(\widetilde{J})
= 2 \widetilde{\mu}(2\widetilde{J}-3)$ and equation \eqref{LinJ} yields
\begin{equation}
      J =
(\widetilde{\mu}/\mu)\widetilde{\lambda}_1\widetilde{\lambda}_2\widetilde{\lambda}_3
      - 3 (\widetilde{\mu}/\mu -1)
      (2 + \widetilde{\lambda}^{-2}_1 \widetilde{\lambda}^{-2}_2)^{-1}.
      \label{Jrocks}
\end{equation}
For the ratios $\widetilde{\rho}_0 / \rho_0$ and $\widetilde{\mu}/\mu$,
we take the values,
\begin{equation}
\widetilde{\rho}_0 / \rho_0 = 1.0, \qquad
\widetilde{\mu}/\mu = 0.6.
\end{equation}

For the principal stretches in the layer we take
\begin{equation}
\widetilde{\lambda}_1 = 1.45, \qquad
\widetilde{\lambda}_2 = 1.05, \qquad
\widetilde{\lambda}_3 = 0.75,
\end{equation}
so that $\widetilde{J}=1.14$, which means a change in volume of 14\% .
The corresponding stress tensor in the layer is
\begin{equation}
      \widetilde{\vec{T}}=\widetilde{\mu} (1.13 \vec{i}\otimes \vec{i} +
       0.25 \vec{j}\otimes \vec{j} - 0.22 \vec{k}\otimes \vec{k}),
\end{equation}
so that the deformation can be maintained with the constant normal pressure
$\vec{\tau}=-0.22\widetilde{\mu} \vec{k}$ (deadload) applied at the
upper face of the layer. It then follows from \eqref{Jrocks} and
\eqref{lambda3}
that $J=1.18$ and
\begin{equation}
\lambda_1 = 1.45, \qquad
\lambda_2 = 1.05, \qquad
\lambda_3 = 0.77
\end{equation}
\begin{figure}
\centering
\epsfig{figure=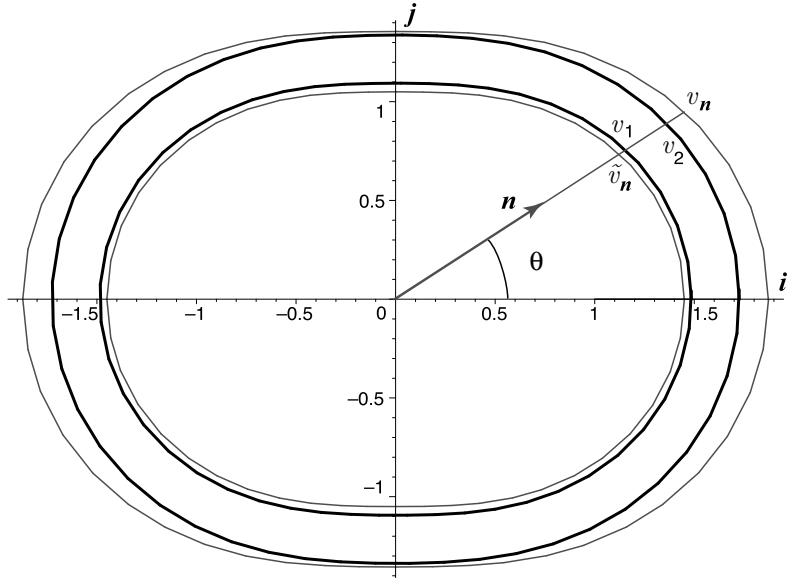, width=.75\textwidth}
   \caption{
   Polar graph of the Love wave speeds $v_1$ and $v_2$ corresponding to
   $kh=\pi$ as a function of the angle
   $\theta$ that the
   propagation direction $\vec{n}$ makes with the direction $\vec{i}$.}
   \label{PolarGraph}
\end{figure}
As explained in Section \ref{section5}, the dispersion curves can be deduced
from dispersion curves in the linear isotropic case, and are shown
in Fig.\ref{dispersioncurves}. Clearly, because $\widetilde{v}_{\vec 
n}$ varies with the
chosen direction $\vec{n}$, this figure shows that the number of 
possible modes for a given
value of the dispersion parameter $kh$ is not necessarily the same 
for all $\vec{n}$.
We here focus on the influence of pre-strain and choose, for 
instance, $kh = \pi$
(wavelength equal to \color{black} twice \color{black} the thickness 
layer), a value
of this parameter such that two modes of propagation are possible in 
all directions
$\vec{n}$~: a \emph{fundamental mode} with speed $v_1$, and a 
\emph{second mode} with
speed $v_2$. In Fig.\ref{PolarGraph}, we plot the polar
graphs of $v_{\vec n}$, $\widetilde{v}_{\vec n}$, and of the Love
wave speeds $v_1$ and $v_2$ of the two modes as
a function of the angle $\theta$ between the propagation direction
$\vec{n}$ and the principal direction $\vec{i}$, corresponding to
the greatest stretch $\widetilde{\lambda}_1 = \lambda_1 = 1.45$. We 
note that the greatest
and least values of $v_1$ and $v_2$ correspond to propagation along 
the directions of
greatest and least stretch in the interface, \color{black}
indicating an experimental way of determining these directions.

\color{black}

%***************************************

\section{Conclusion}
\label{section8}

%***************************************

In this paper, we have obtained an exact finite-amplitude Love wave 
solution for a layer
and a subtrate consisting of pre-strained compressible neo-Hookean 
materials. The
dispersion relation is similar to that of linear isotropic 
elasticity, but an explicit
dependence on the pre-strain is exhibited. In particular, the number 
of wave modes for
different values of the dispersion parameter $kh$ ($k$: wave number, 
$h$: thickness of the
deformed layer) is influenced by the pre-strain.

It should be \color{black} emphasized \color{black}
that the existence of this Love wave 
solution is subjected to the
condition that the propagation direction $\vec{n}$ and the normal 
$\vec{m}$ to the interface
are such that 
$\vec{n}\vec{B}\vec{m}=\vec{n}\widetilde{\vec{B}}\vec{m}=0$, where 
$\vec{B}$
and $\widetilde{\vec{B}}$ are the left Cauchy-Green strain tensors 
characterizing the
pre-strain of the subtrate and of the layer, respectively. Note that 
it follows from the
continuity of the traction at the interface that the conditions 
$\vec{n}\vec{B}\vec{m}=0$
and $\vec{n}\widetilde{\vec{B}}\vec{m}=0$ are equivalent.

Thus, the interface may be arbitrarily chosen. However, when it is 
not a principal plane of
$\vec{B}$ and $\widetilde{\vec{B}}$, there is only one propagation 
\color{black} direction \color{black} satisfying these
conditions. 
\color{black} In contrast\color{black}, 
when the interface is a principal plane 
of $\vec{B}$
and $\widetilde{\vec{B}}$, all propagation directions $\vec{n}$ in 
this interface are
possible. In this case, for a given value of $kh$, the number of 
possible modes is not
necessarily the same for all propagation directions.

The energy flux and energy density of the solution in the layer and 
in the substrate have
been studied in detail. In particular, it has been shown that the 
mean energy fluxes in
the layer and in the substrate are both along 
$\vec{B}\vec{n}=\widetilde{\vec{B}}\vec{n}$,
thus along the same direction, parallel to the interface. The fact 
that this direction is
not along the propagation direction (except when $\vec{n}$ is a 
principal direction) is due
to the anisotropy induced by the pre-strain. \color{black}

%**************************

%***********************

%************************************

\appendix

%************************************

\section{Uniqueness for the determination of $J$ in the substrate}

Here we consider the equation \eqref{eq:detJ} for the detemination of 
$J$ in the
substrate when the strain tensor $\widetilde{\vec B}$ in the layer is 
given. It may also be
written as
\begin{equation}
     f(J)\equiv \demi G'(J) + \mu J \widetilde{J}^{-2}\vec{m} \cdot 
\widetilde{\vec{B}} \vec{m} =
     \vec{m}\cdot \widetilde{\vec T} \vec{m},
     \label{eq:A.1}
\end{equation}
where $\widetilde{\vec T}$ is the stress tensor in the layer, corresponding to
the strain tensor $\widetilde{\vec B}$.

First, using the strong ellipticity conditions \eqref{S-E}, we note that
\begin{equation}
     f'(J) = \demi G''(J) +
     \mu \widetilde{J}^{-2}\vec{m} \cdot \widetilde{\vec{B}} \vec{m} >0,
     \label{eq:A.2}
\end{equation}
so that $f(J)$ is strictly monotonous increasing for $J \in [0, \infty]$.
Then, recalling $G'(1)=-2\mu$, we have $\demi G'(J) \leq -\mu$ for 
$J \leq 1$,
and $\demi G'(J) \geq -\mu$ for $J \geq 1$,  hence
\begin{equation}
     \lim_{J \rightarrow 0}f(J)= \demi \displaystyle \lim_{J 
\rightarrow 0} G'(J) \leq -\mu, \qquad
     \lim_{J \rightarrow \infty}f(J)= \infty.
     \label{eq:A.3}
\end{equation}
Owing to the monotonicity of $f(J)$, the limit for $J \rightarrow 
0$ exists and is either
finite and negative or $-\infty$.

If $\demi \lim_{J \rightarrow 0} G'(J)=-T_0\leq -\mu <0$ (with 
$T_0$ finite), then, clearly,
for any strain tensor $\widetilde{\vec B}$ such that $\vec{m}\cdot 
\widetilde{\vec T}
\vec{m}>-T_0$, equation \eqref{eq:A.1} has exactly one solution for 
$J >0$. However, for any
strain tensor $\widetilde{\vec B}$ such that $\vec{m}\cdot \widetilde{\vec T}
\vec{m}\leq -T_0$, this equation has no solution for $J >0$.

If $\lim_{J \rightarrow 0} G'(J)=-\infty$, then, clearly,
whatever be the value of $\vec{m}\cdot \widetilde{\vec T} \vec{m}$, 
equation \eqref{eq:A.1}
has exactly one solution for $J >0$.

\end{document}